\documentclass[11pt]{cernrep}
\usepackage{graphicx}
\usepackage{here}
\begin{document}
\title{Novel  QCD phenomena in pA collisions at LHC}
\author{L.~Frankfurt$^1$,  M.~Strikman$^2$}
\institute{Nuclear Physics Dept.,
Tel Aviv University, Israel$^1$,
Department of Physics, PSU, USA$^2$}

\maketitle
\begin{abstract}
We discuss novel QCD phenomena in pA collisions at LHC energies such 
as the possibility  to invesigate the behaviour of hard processes at 
$x $ as low as $10^{-7}$, signals for onset of the black body limit
 in the hard processes and in the nucleon fragmentation, multijet production.

\end{abstract}

\section{Physics motivations}
\label{aintro}
The proton-ion collisions at LHC will be qualitatively different from those 
at fixed target energies \footnote{An extensive discussion of the novel pA 
physics at LHC was presented in \cite{FELIX}. Here we summarize few
key points of this analysis and extend it to consider effects of the
small x high field regime.}   In the soft interactions two prime effects are 
a strong increase  of the mean number of ``wounded'' nucleons - from about 6 
to about 20 in $pA$ central collisions (due to increase of 
$\sigma_{inel}(NN)$) and a 50 $\%$ increase of the average impact parameters 
in $pp$ collisions (as manifested in the shrinkage with increase with energy 
of the diffractive peak in the elastic $pp$ collisions)
making it possible  for a  proton to interact  often  simultaneously 
with nucleons which have essentially the same longitudinal coordinate but 
different impact parameters. At the same time geometry of $pA$ collisions at 
LHC should be more close to that in the classical mechanics because increase 
of cross sections combined with the suppression of the inelastic diffraction 
in $pp$ scattering, leads to the suppression of fluctuations in the value of 
effective $NN$ cross section (and hence the suppression of inelastic shadowing
corrections in $pA$ scattering), to a  very strong suppression of inelastic 
diffraction in $pA$ collisions as compared to the fixed target FNAL
energies \cite{FMS}. In the hard interactions the prime effect is a 
strong increase of the gluon densities at small $x$ both in projectile and 
ion. Gluon densities can be measured at LHC up to ultra small $x$ 
$\propto 10^{-6 \div 7}$  where PQCD will be probed in a new domain
\cite{FELIX}. 
As a consequence of blackening of the 
interaction of proton partons with nuclear partons at small $x_A$ 
one should expect disappearance in the proton fragmentation region of hadrons 
with small transverse momenta, and hence a strong suppression of 
nonperturbative QCD effects
\footnote{A similar phenomenon of the suppression of small transverse momenta 
for another quantity, the nuclear light cone wave function 
at small $x$,  is  discussed in 
the saturation models \cite{Muellerrev,Iancu:2002xk}.}
Thus the high transverse momentum proton fragmentation region is a natural 
place to search for unusual QCD phases which should live long time because 
of the Lorentz dilatation of time. In spite of a  small coupling constant, 
gluon densities of heavy ions become larger in a wide range of $x$ and $Q^2$ 
and impact parameters than that permitted by the conservation of probability 
within the leading twist approximation 
\cite{FKS96,FELIX,Muellerrev,Iancu:2002xk,FGSrev}. 
Thus 
a decrease of the strong coupling constant with virtuality  
would be insufficient to support applicability of PQCD to the hard small $x$ 
processes in the kinematical domain where gluon fields may appear 
strong enough 
for nonperturbative QCD vacuum to become unstable  and where equations 
of QCD may need to be modified. The possibility for novel QCD regime and 
therefore new phenomena are maximized in this QCD regime. \footnote{It is 
important 
to distinguish between parton densities which are defined within the leading 
twist approximation only \cite{Feynman} and may increase with energy forever 
and physical quantities - cross sections/structure functions - whose increase 
with energy at given impact parameter is restricted by the probability 
conservation.} At the same time interpretation of these novel effects would 
be much more definitive in $pA$ collisions than in the heavy ion collisions.

New classes of strong-interaction phenomena which occur within the 
short-distance, near-the-light-cone space-time region and which could be  
associated with perturbative QCD or with interface of perturbative and 
nonperturbative QCD are much  more probable in $pA$ collisions than in $pp$ 
collisions. The densities of partons, and of energy-momentum, may be high 
enough to ``burn away'' the nonperturbative QCD vacuum structure in a cylinder 
of a radius $\sim 1 fm$ and length $\sim 2 R_A$, leaving behind only a dense 
partonic fluid governed by new QCD dynamics with small $\alpha_s$ but 
effective short-distance large
couplings. If there is a large enhancement of hard-collision, gluon-induced 
processes, then there should be an enhancement of hard multiparton collisions,
heavy-flavor production  \cite{FELIX}. The study of the $x$, $p_t^2$ and 
$A$-dependence of multiparton interactions, of the forward charm and beauty 
hadron production (in the direction of the fragmenting proton) should be 
especially incisive to address these questions.

Estimates of the kinematical region for the new QCD dynamics in proton 
(nucleus)-heavy nucleus collisions based on the LO formulae for the dipole 
cross section and the conservation of probability indicate \cite{FGSrev} 
that for the central impact parameters for $A\sim 200$ and $Q^2\sim 25 GeV^2$
for quarks $x_A\leq 5\cdot 10^{-5}$ and  for gluons $x_A\leq  10^{-3}$.

\section{Measurements of  nuclear parton densities at ultra small x}

\noindent
One of the fundamental issues in high energy QCD is the dynamics of the hard 
interactions in the small $x$ kinematics. Depending on the resolution scale 
($Q^2$) one investigates here either the leading twist effects or the regime 
of strong color fields. Since the cross-sections for hard processes increase 
roughly linearly with $A$
(except for very small $x$ and relatively small  $Q^2$ where the counting
rates are high anyway), even short runs with nuclear beams will produce data 
samples sufficient to measure parton distributions with {\it statistical } 
accuracy better than 1$\%$  practically down to the smallest $x$ which is 
kinematically allowed: $x \sim 10^{-7} $ for quarks and $x \sim 10^{-6} $  
for gluons \cite{ACSW}. The systematic errors for the ratios of nuclear and 
nucleon structure functions are also expected to be small since most of these
errors cancel in the ratios of the cross-sections, cf \cite{HERAnuc1}.

There are several processes which could be used to probe the small
$ x$ dynamics in $pA$ collisions. They include the Drell-Yan pair production,
study of dijets, jet + photon, charm production, diffractive 
exclusive production of three jets \cite{3jet}, multijet events.
It is worth emphasizing that for these measurements it is  necessary to 
detect jets, leptons and photons at rapidities close enough to the nucleon 
rapidity. In  this $y$-range, the  accompanying soft hadron multiplicities 
are relatively small, leading to a soft particle background comparable to 
that in $pp$ scattering.  In fact, for $|y_{max}-y| \leq (2-4)$ the 
background level is likely to be significantly  smaller than in $pp$ 
collisions due to suppression of the leading particle production in $pA$ 
scattering (for a thorough discussion of these reactions see \cite{FELIX}).

The measurements of the dimuon production appear to be more feasible  
experimentally  for the forward angles than studies of the hadron  production 
due to small energy losses for  muons. In the kinematics where the leading 
twist dominates one can study both the quark small $x$ distributions  (from 
the cross section at  relatively
small transverse  momenta:
$p_t^2(\mu^+\mu^-) \ll  M^2(\mu^+ \mu^-)$, and the gluon distribution 
(from the measurements of the cross section at 
$p_t(\mu^+\mu^-) \sim M(\mu^+\mu^-)$ \cite{Catani:2000jh}). 
Note also that for the forward region a dimuon background from the charm 
production is likely to be rather small since muons in average carry
only 1/3 of the energy in the charm decays.

\section{Signals for onset of the black body limit}
\label{sec3}

A fast increase with energy of structure functions of heavy nuclei contradicts 
to the probability conservation for a wide range of parton momenta and 
virtualities for the LHC energy range. On the contrary the rapid, power-like 
growth of structure function of a nucleon observed at HERA may continue 
forever because of the diffuse edge of a nucleon \cite{bbl}.  At present
new  QCD dynamics of the regime of high parton densities where multiparton 
interactions are inhibited remains a subject of 
lively debates. PreQCD 
black body regime for the structure functions of heavy nuclei \cite{Gribov}, 
QCD black body regime for hard processes off heavy nuclei \cite{bbl}, 
saturation of parton densities at given impact parameter \cite{Muellerrev}, 
color glass condensate regime \cite{Iancu:2002xk}, turbulence and related 
scaling laws are possible options. For the estimates of the pattern of the 
novel phenomena we shall use in this text the formulae of black body regime 
because  they are generic enough  for the expectations of the
onset of new QCD dynamics within the existing theoretical approaches.
\footnote{In the  
black regime highly virtual probe interacts at the same time with  several 
high 
$k_t$ partons from the wave function of a hadron \cite{bbl}.
This is qualitatively different from the asymptotic freedom limit 
where interaction with only one highly virtual parton are important.
Hence the 
interacting 
parton is far from being free when  probed by a
hard probe of a large  coherence length 
$l_c=1/2m_Nx$. }

The use of nuclei because of their uniform density and a possibility to 
select scattering at the central impact parameters allows one to enhance the 
high parton density effects. Indeed, in the small $x$  regime a hard collision
of  a parton of the projectile nucleon with a parton  of the nucleus occurs 
coherently with all the nucleons at a given impact parameter. The coherence 
length $\approx 1/2m_{N}x$ by far exceeds the nuclear size: in the kinematic 
regime accessible at LHC it can reach up to $10^5$ fm (in the nucleus rest 
frame). In the rest frame of the nucleus this can be visualized in terms of  
the propagation of a parton  in high density gluon fields over much larger 
distances than is possible with free nucleons. In the Breit frame it 
corresponds to the fact that small $x$ partons cannot be localized 
longitudinally to anything close to the Lorentz-contracted thickness of the  
nucleus.  Thus low $x$  partons from different nucleons overlap  spatially, 
creating much larger parton densities than in the free nucleon case. This 
leads to a large amplification of the multiparton hard collisions expected in 
QCD at small $x$; see e.g. \cite{HERAnuc1,FELIX}. Eikonal approximation 
\cite{MQ}, constraints from the probability conservation 
\cite{FKS96,AFS} indicate that new QCD dynamics should reveal itself at 
significantly larger $x$ than in the proton and in more striking effects.
Recently extensive studies of hard diffraction were performed at HERA, 
for review see \cite{Abr2000}. This allows to evaluate 
taming due to the leading twist nuclear shadowing which turns out to be 
large both in the gluon and quark channels and significant up to very large 
$Q^2$, - for the recent discussion see \cite{FGSrev}. However the leading twist taming being large is 
still insufficient to restore the probability conservation
leaving a lot of room for the violation of leading twist 
approximation. It maybe more serious than merely
due to enhancement of the high twist effects as the whole picture of 
interactions can qualitatively change. Really, the common wisdom is that 
smallness of $\alpha_s$ guarantees that bare particle is free within a hadron 
but this remarkable property which confirmed experimentally in many hard 
processes is violated in small x phenomena. One may   think about this 
phenomenon in term of an effective violation of the asymptotic freedom in 
this limit. It is expected to occur 
in heavy nuclei at $x$  at least ten times larger than in the 
proton. Hence in the LHC kinematics the region where novel QCD phenomena are 
important  will extend at least two  orders of magnitude in $x_A$
for a large range of virtualities.

\noindent
{\bf 
The space time picture of the black body limit in the proton-nucleus 
collisions}
can be seen best in the rest frame for the nucleus. A parton belonging 
to the proton emits a hard gluon (virtual photon) long before the target
and interacts with the target in a black regime releasing the fluctuation,
e.g. a Drell-Yan pair.  This leads to a qualitative change in the
picture of the proton-nucleus interactions for the partons with 
$x_p, p_t$ satisfying the condition that 
\begin{equation}
x_A={4 p_t^2\over x_{p}s_{NN}}
\label{limit}
\end{equation}
is in the black body kinematics for the resolution scale
$p_t\leq p_t^{b.b.l.}(x_A)$. Here $ p_t^{b.b.l.}(x_A)$ is maximum $p_t$ for 
which the black body approximation is applicable. 
In the kinematics of LHC $Q^2\approx 4(p_t^{b.b.l.})^2$ can be estimated 
by using formulae derived in  \cite{FKS96}.  At minimal $x_A$ it may appear as 
large as $(p_t^{b.b.l.}(x_A))^2=15 GeV^2$ .
All the partons with such $x_p$ will obtain $p_t(jet)\sim  p_t^{b.b.l.}(x_A)$ 
leading to the multijet production. The black body regime will extend down in 
$x_p$ with increase of the incident energy.  For LHC for $p_t \leq 3 GeV/c$ 
this regime may  cover the whole region of $x_p \geq 0.01$ where of the order 
ten partons resides. Hence in this limit most of the final states will 
correspond to multiparton collisions.
For  $p_t \leq 2 GeV/c$ the region extends to $x_p \geq 0.001$. At LHC collisions
 of such partons correspond to the central rapidities.
Dynamics of conversion of the high $p_t$ partons with similar rapidities to
hadrons is certainly a collective effect which 
requires a special consideration.

{\bf Inclusive observables}
For the calculation of the total cross sections the logic similar to 
the one used in the consideration of $\gamma^*$ - nucleus
scattering \cite{bbl} should be applicable. In particular for the case of 
the total cross section of the dimuon production 
can be evaluated as:
\begin{equation}
{d\sigma(p+A \to \mu^+\mu^- + X)\over dx_Adx_p}=
{4\pi\alpha^2\over 9}{K(x_A,x_p,M^2)\over M^2}
F_{2p}(x_p,Q^2)\cdot {1\over 6\pi^2} M^2\cdot 2\pi R_A^2 \ln(x_0/x_A),
\end{equation}
for large but not too large dimuon masses, $M^2$. In particular we
estimate 
 $M^2(bbl)(x_A=10^{-7}\approx 60 GeV^2$.
Here $K$-factor
has the same origin as in the leading twist case, but it should be smaller 
since it originates from the gluon emissions only from the 
parton belonging to the proton, $R_A$ is the nuclear radius, and $x_0$
is maximal $x_0$ for which the black body limit is valid.
Hence the expected $M^2$ dependence of the dimuon production is
qualitatively different in this case than in the case of $pp$ scattering
where scattering at large impact parameters may mask the black body
limit contribution. Also the $x_A$ dependence in this limit becomes
pretty weak.

The study of the $p_t$ distribution of the dimuons may provide also a signal 
for  the onset of the black body regime. Similar to the case of $p_t$ 
distribution of leading partons in the deep inelastic scattering \cite{bbl} 
one should expect a broadening of the $p_t$ distribution of the dimuons in 
the black body limit as compared to the DGLAP expectations, see \cite{Jamal} 
for a calculation of this effect in  the color glass condensate model.

With further decrease of $x_A$ formulae of black limit 
will probably overestimate cross section because the interaction  with heavy 
nucleus of sea quarks and gluons in the proton would become black also.  

The onset of the black body limit will lead also to gross changes in the
hadron production: much stronger drop with $x_F$ of the spectrum in the 
proton fragmentation region accompanied by a gross $p_t$ broadening of the 
spectrum as well as the enhancement of the hadron production at smaller 
rapidities, see discussion in section \ref{a.4.4}.

It is worth emphasizing here that onset of the black body limit for ultra
small $x_A$ will pose limits on using the forward kinematics for the 
measurements of the  parton densities at larger $x_A$. Indeed as soon as one 
tries to use large enough $x_p$ for the nuclear parton density measurements
one would have to take into account that collisions of these partons with 
nuclear partons are always accompanied by a significant $p_t$ broadening 
associated with the  interaction with ``black component'' of the nucleus 
wave function. As a minimum this would lead to a significant 
$p_t$ broadening of the $p_t$ distribution of the produced hard system. 
One can also question the validity of the leading twist expansion  for the 
cross sections integrated over the transverse momentum of the produced system.
To investigate these phenomena an ability to do measurements at fixed $x_A$
and different $x_p$ would be very important.

\section{Mapping of the three dimensional  nucleon  parton structure}

The systematic studies of hard inclusive processes during the last two 
decades have led to a pretty good understanding of the single parton densities
in nucleons. However very little is known about multiparton correlations in 
nucleons which can provide critical new insights into the dynamics of the 
strong interactions, and allow to discriminate between different models of 
nucleons. Such correlations may be generated, for example, by the fluctuations
of the transverse size of the color field in the nucleon leading, via color 
screening, to correlated fluctuations of the densities of gluons and quarks.

A related source of correlations is QCD evolution, since a selection of a 
parton with a given $x, Q^2$ may lead to a local (in transverse plane) 
enhancement of the parton density at different $x$ values.
Also, practically nothing is known about  possible  correlations 
between the transverse size of a particular configuration in the 
nucleon and the longitudinal distribution of partons in this configuration.

\subsection{Multi-jet production and double parton distributions}
It was recognized already more than two decades ago \cite{dpth} that 
the increase of parton densities at small $x$ leads to a strong increase of 
the probability of nucleon-nucleon collisions where two or more partons of 
each projectile experience pair-vice independent hard interactions.
Although the production of multijets through the double parton
scattering mechanism was investigated in several experiments
\cite{dpex,cdf} at $pp,p\bar p$ colliders, the
interpretation of the data was hampered by the need to model both
the longitudinal and the transverse partonic correlations at the
same time. The studies  of  proton-nucleus collisions at LHC will provide a
feasible opportunity to study separately the longitudinal and transverse 
partonic correlations in the nucleon as well as to check the validity of 
the underlying picture of multiple collisions.

The simplest case of a multiparton process is the double parton collision. 
Since the momentum scale $p_t$ of a hard interaction corresponds 
to much smaller  transverse distances $\sim 1/p_t$ in the
coordinate space than the hadronic radius, in a double parton
collision the two interaction regions are well separated in the
transverse space. Also in the c.m. frame pairs of
partons from the colliding hadrons are located in pancakes of
thickness $\leq (1/x_1 +1 /x_2)/p_{c.m.}$. So two hard collisions
occur practically simultaneously as soon as $x_1, x_2$ are not too
small and hence there is no cross talk between two hard collisions.
A consequence is that  the different parton
processes add incoherently in the cross section. The double parton
scattering cross section, being proportional
to the square of the elementary parton-parton cross section, is
therefore characterized by a scale factor with dimension of the
inverse of a length squared. The dimensional quantity is provided
by the nonperturbative input to the process, namely by the
multiparton distributions. In fact, because of the localization of
the interactions in transverse space, the two pairs of colliding
partons are aligned, in such a way that the transverse distance
between the interacting partons of the target hadron is
practically the same as the transverse distance between
the partons of the projectile. The double parton distribution is
therefore a function of two momentum fractions
and of their transverse distance, and it can be written as
$\Gamma(x,x',b)$. Actually $\Gamma$ depends also on the
virtualities of the partons, $Q^2,Q'^2
 $, though to make the expressions more compact we will
not write explicitly this $Q^2$ dependence. Hence the double
parton scattering cross section for the two ``two $\to$ two''
parton processes $\alpha$
 and  $\beta$ in an inelastic interaction between hadrons
$a$ and $b$ can be written as:
\begin{eqnarray}
\sigma_D(\alpha,\beta)&=&{m\over2}\int\Gamma_a(x_1,x_2;b)\hat{\sigma}_{\alpha}
(x_1,x_1')
\hat{\sigma}_{\beta}(x_2,x_2')\Gamma_b(x_1',x_2';b)dx_1dx_1'dx_2dx_2'd^2b
\label{1}
\end{eqnarray}
\par\noindent
where $m=1$ for indistinguishable parton processes and $m=2$ for
distinguishable parton processes. Note that though the factorization 
approximation of Eq.(\ref{1}) is generally accepted in the analyses of the 
multijet processes and appears natural based on the geometry of the process 
no formal proof exists in the literature. As we will show below the study of 
the A-dependence of this process will allow to perform  a stringent test of
this  approximation.

Here to simplify the discussion we
neglect small non-additive effects in the parton densities, which
is a reasonable approximation for $0.02\leq x \leq 0.5$. In this
case we have to take into account only $b$- space correlations of
partons in individual nucleons.

 One has therefore two different contributions to the double parton 
scattering cross section: $\sigma_D=\sigma^1_D+\sigma^2_D$. The first one,
$\sigma_1^D$, interaction with two partons of the same nucleon in the nucleus,
is the same as for the nucleon target (the only
difference being the enhancement of the parton flux) and the
corresponding cross section is
\begin{equation}
\sigma^1_D=\sigma_D\int d^2BT(B)= A\sigma_D,
\label{sigma1}
\end{equation}
\par\noindent
where
\begin{equation}
T(B)=\int_{-\infty}^{\infty} dz \rho_A(r), \int T(B)d^2B=A
\end{equation}
\label{TB}
\par\noindent
is the nuclear thickness,
as a function of the impact parameter of the
 hadron-nucleus collision $B$.

The contribution to the term in $\Gamma_A(x_1',x_2',b)$ due to the
partons originated from different nucleons of the target, $\sigma^2_D$,
can be calculated {\it solely} from the geometry of the process by
observing that the nuclear density does not change within a
transverse scale  $\left<b\right> \ll R_A$. It rapidly increases with A   
$\propto \int T^2(B)d^2B$. Using information from the CDF double scattering 
experiment\cite{cdf} on the mean transverse separation
of partons in a nucleon,  one finds that   the contribution of the
second term should dominate in the case of proton - heavy nucleus collisions:
 $\sigma^1_D/\sigma^2_D\approx .68 (A/12)^{.39}$ \cite{ST}. 
Hence one expects a stronger than $\propto A$ increase of the multijet
production  in $pA$ collisions at LHC. Measurements with a set of nuclei
would  to measure the double parton distributions in nucleons and also to
check  the validity of the QCD factorization for such processes which 
appears natural but which so far was not derived in pQCD.  An important 
application of the discussed process would be to investigate  transverse
correlations between the nuclear partons in the shadowing  region. This
would require a selection of   both partons from the nucleus in the 
shadowing region, $x_A\leq  x_{sh} \sim 10^{-2}$. \footnote{The A-dependence
of the ratio of $\sigma_2/\sigma_1$ in the kinematics where only one of
the nuclear partons has $ x_A\leq x_{sh}$ is practically the same as for the 
case when both nuclear partons have $x\geq x_{sh}$.}

As we already mentioned in section \ref{sec3} partons with 
sufficiently 
large $x_p$ satisfying Eq.\ref{limit} are expected to interact
with a probability of the order one with small $x$ partons of nucleus.
As a result  a hard collision of the partons with sufficiently large $x_p$ and 
$x_A\geq 0.01 $ will be accompanied by production of a minijet/minijets at 
the black body kinematics with $p_t \sim  p_t^{b.b.l.}(x_A)$. Most of 
sufficiently fast partons of the protons will generate minijets leading to 
a strong suppression of cross section of the events with only two jets.

Detectors with sufficiently large  rapidity acceptance  would allow to  
detect events originating from the triple parton collisions
at $x_A\geq x_{sh}$ and large enough $p_t$ where $p_t$ broadening related to
the blackbody limit effects are sufficiently small. This provide  stringent 
tests of the dynamics of the hard interactions and provide  additional 
information on two parton correlations as well as unique information on 
triple parton correlations.

Other opportunities with multi-jets include
\begin{itemize}
\item Probing correlations between partons in the nucleus at high
densities, i.e. for $x_{1~A}, x_{2~A} \ll 10^{-3}$, which would
provide qualitatively new information about the dynamics of nuclear
shadowing   and  the presence of possible new condensates of partons.
\item Studying the accompanying  soft hadron production (cf. the discussion in 
Section~\ref{a.1.5}), which would allow measurement of
the transverse  size of a proton configuration containing
partons $x_{1}, x_{2}$. In particular, the production of  
4 jets with  two jets at large $x$ in the proton
fragmentation region $x_{1}+x_{2} \geq 0.5$ may provide another way to 
look for point-like configurations in nucleons, see  \cite{3jet}
and \cite{FELIX} for discussion of some other options.
\end{itemize}

\subsection{Proton-ion collisions probe  transverse nucleon structure}
\label{a.1.5}

As we discussed in \cite{3jet}, one supposes that in some subclass of 
events the distribution of constituents in the initial proton may be 
unusually local in the transverse (impact 
parameter)
plane when the proton collides with the ion. If this is so, its effective 
cross-section per nucleon will be greatly reduced, perhaps all the way to the 
perturbative-QCD level. If the effective cross-section of such a point-like 
configuration goes below 20mb, there will be an
appreciable probability that it can penetrate through the center
of a $Pb$ ion and survive. This would lead to a highly enhanced yield of
diffractive production of the products of the point-like configuration in
collisions with heavy ions.

Not only might the properties of the final-state collision products depend
upon the nature of the transverse structure of the proton primary at
arrival at the collision point, but even the conventional parton
distributions may also be affected. For example, let us  consider, a nucleon 
as a 
quark and small diquark connected by a
narrow QCD flux-tube. It should be clear that if the flux-tube is at right
angles to the collision axis at arrival, then the valence partons will
have comparable longitudinal momentum or $x$. On the other hand, if the flux 
tube is parallel to the direction of motion, then one of the valence systems 
will have very large $x$, and the other very small. This happens because in 
this case the internal longitudinal momenta of the valence
systems, in the rest frame of the projectile proton, are in opposite
directions. Therefore the smallness of the configuration is correlated
with the joint $x$-distribution of its constituents. This kind of
nonfactorization may be determined by the study of the perturbative-QCD
processes of dilepton, direct-photon, or dijet production as a function of
the centrality of the collision, multiplicity of soft hadrons, as
 well as   a function of the atomic number.
\footnote{The  impact-parameter dependence is accessible in the 
proton - nucleus collisions via a  study of  nuclear  fragmentation into a 
number of channels, for an extensive discussion see \cite{FELIX}.}
Naturally such studies would require also experimental investigation
of the production of the soft hadrons as a function of impact
parameter at LHC energies where it may differ quite strongly 
from the one observed at fixed target energies.

One possible kinematics where a strong correlation is expected
is when a parton with large $x$($x\geq 0.6$)  is selected  in 
the proton. The presence of such a parton  requires three quarks
to exchange rather large momenta. Hence  one may expect that these
configurations have a smaller transverse size and therefore interact with 
the target with a smaller effective
cross-section $\sigma_{eff}(x)$. Suggestions for such a
dependence of the size on $x$ are widely discussed in the literature.
Using as a guide a geometric (eikonal type) picture of $pA$ interactions 
and neglecting (for simplicity) shadowing effects for nuclear parton densities
one can estimate the number of wounded nucleons $\nu(x,A)$ in events
with a hard trigger (Drell-Yan pair, $\gamma$-jet, dijet,...)
as a function of $\sigma_{eff}$ \cite{FS85}:
\begin{equation}
\nu (x,A)= 1 + \sigma_{eff}(x) 
{A-1 \over A^2}\int T^2(b)~d^2b,
\label{nu}
\end{equation}
where the nuclear density per unit area $T(b)$ is defined in Eq.(~\ref{TB}).
At LHC for average inelastic $p Pb$ collisions and $\sigma_{eff} \sim
\sigma_{inel}(pp)$ Eq.\ref{nu} leads to    $\nu \approx $ 10. This is somewhat
larger than the average number of wounded nucleons in $pA$ collisions, due to 
the selection of more central impact parameters in events with a hard trigger.

A decrease of the effective cross-section for large $x$, say, by a factor of 
2,   would result in a comparable drop of the number of particles produced at 
central rapidities as well as in a smaller number of nucleons produced in the 
fragmentation of the nucleus. in the nucleus fragmentation region.

\subsection{$A$-dependence of the particle production}
\label{a.4.4}

\paragraph*{Particle production in the proton fragmentation region}

\noindent
The $A$-dependence of hadron production in the proton fragmentation region
remains one of the least understood aspects of hadron-nucleus interactions. 
Practically all  available data are inclusive and correspond to the energies 
where the cross section of inelastic $NN$ scattering is about three times 
smaller than at LHC. They  indicate that the cross-section is dominated by 
the production of leading particles at large impact parameters, 
where the projectile interacts with only one or two nucleons of the
nucleus. As a result very little information is available about hadron
production at the central impact parameters which are most crucial for the 
study, for example of $AA$ collisions. Theoretical predictions for this 
region are also rather uncertain.

In  eikonal type models, where the energy is split between several soft 
interactions, one may expect a very strong decrease of the yield of the leading particles.
The dependence is expected to be exponential with path length in the nucleus; 
their mean energy is attenuated exponentially. On the other hand, if the 
valence partons of the projectile do not lose a significant amount of their 
initial momentum, as suggested by the models motivated by perturbative QCD, 
see review in \cite{Baier}, the spectrum of leading particles may approach a 
finite limit for large $A$ and  central impact parameters
\cite{Ber}. Indeed in this case the leading partons will acquire significant
 transverse momenta and will not be able to coalescence back into leading 
baryons and mesons as it seemingly happens in the nucleon-nucleon collisions. 
As a result they will fragment practically independently leading to much 
softer distributions in the longitudinal momentum for mesons and
especially for baryons \cite{Ber}.

We discussed above that the strength of the interaction of fast partons  in 
the nucleon ($x_p \geq 10^{-2}$ for LHC) with heavy nuclei at the central 
impact parameters may approach the black body limit and that under this 
scenario all these partons will acquire large transverse momenta 
$\sim p_t^{b.b.l.}(x_A)$.
\footnote{Hence in the black body limit $p_t$ broadening of the partons
 should be much larger than at low energies where $p_t$ broadening is
consistent with the QCD multiple rescattering model \cite{Baier},  
see \cite{DFS} for the discussion of the matching of these two regimes.}
This would lead  to
a very significant $p_t$ broadening of the 
spectrum of leading hadrons \cite{Jamal1}. Probably the most feasible 
way to study this effect will be to measure the $p_t$ distribution of the 
leading neutrons for the collisions at the central impact parameters, 
which will closely follow the $p_t$ distribution of leading quarks \cite{JDS}
It is worth emphasizing that a strong $p_t$ broadening which should
increase with increase of $x_F$ clearly distinguishes this mechanism 
of the suppression of the leading hadron spectrum from the soft physics 
effect of the  increase of the total $pp$ cross-section by nearly a factor 
of three from fixed target energies.

\paragraph*{Particle production in the central region}
\noindent
The Gribov-Glauber model including inelastic screening effects, but neglecting
the final state interactions between the hadrons produced in different Pomeron
exchanges, as well as neglecting the interactions of these hadrons with 
the nuclear target, leads to a  prediction that at high energies and 
rapidities $y $, such that $|y_p -y| \gg 1$, the inclusive spectrum of 
produced hadrons should be proportional to that in the inclusive 
cross-section of $hN$ scattering:
\begin{equation}
{d \sigma^{p+A \to h+X}(y, p_T)\over dy~ d^2p_T}= A  
 {d \sigma^{p+N \to h+X}(y, p_T)\over dy~ d^2p_T}.
\label{a13}
\end{equation}
Since $\sigma_{inel}(pA)/\sigma_{inel}(pN) \propto A^{2/3}$,
Eq.(~\ref{a13}) implies that the multiplicity of particles produced in
the inelastic $pA$ collisions should increase with $A$, roughly as
$A^{1/3}$. Data at fixed target energies do not contradict this
relation, but the energy is too low for an unambiguous interpretation.
Effects of splitting the energy between multiple soft $NN$ interactions
are much larger in the nucleus-nucleus collisions than in
proton-nucleus collisions so checking whether similar formula is valid
for AA collisions would require energies much higher than those
available at RHIC.

The opposite extreme is to assume that interactions of partons in the
black body regime become important in a large range of $x_p\geq
10^{-2}$.  In this case multiple scatterings between the partons,
their independent fragmentation together with associated QCD radiation
(gluon bremsstrahlung) may produce the major part of the total entropy
and transverse energy and will lead to enhancement of the production
of hadrons at central rapidities as compared to the the expectations
of the Gribov-Glauber model.

\paragraph*{Particle production in the ion fragmentation region}
\noindent
For the rapidities close to the nucleus rapidity, there are indications of the 
contribution of slow hadron re-interactions, which lead to an increase  of 
the multiplicity of  nucleons with momenta $1.0 \geq p_N \geq 0.3$ GeV, as 
compared with expectations neglecting final state re-interactions (a factor 
of $\sim 2$ for heavy nuclei) \cite{FS81}. As we have discussed above, it is
doubtful that the Gribov-Glauber picture is correct at LHC energies, 
so one may expect significant deviations from the expectations based on 
Eq.(~\ref{a13}). 
It is likely that color quark system with rapidities close to $y_A$
will be excited along the cylinder
of the radius $\sim$ 1 fm and length $R_A$. This may result in unusual properties
of the hadron production in the nucleus fragmentation region including production of exotic
multiquark, multigluon\ldots  states.

It must be emphasized that a quantitative understanding of the physics of particle
production in $pA$ collisions in this phase-space region is a prerequisite
for the understanding the corresponding phenomenology in $AA$ collisions.

In conclusion, studies of proton-nucleus scattering  at LHC will allow to study
strong interactions in the high field domain over extended region 
of rapidities.
New phenomena will be especially prominent in the proton fragmentation 
region, though
they will also extend to the central region and to 
the nucleus fragmentation region.

It is our pleasure to thank J.Bjorken for the joint study of the $pA$
 physics for the FELIX project and numerous discussions of the
 forward physics. We also thank  GIF and DOE for support.

\end{document}